\documentstyle[preprint,prb,aps]{revtex}
\begin{document}
\draft
\title{Phase diagrams of the $\mbox{\boldmath $S$} = \frac{1}{2}$ quantum 
antiferromagnetic $XY$ model on the triangular lattice
in magnetic fields}
\author{N. Suzuki and F. Matsubara}
\address{Department of Applied Physics,
Faculty of Engineering, Tohoku University,
Sendai 980-77, Japan}
\date{December 10, 1996}
\maketitle

\begin{abstract}
We study the $S = \frac{1}{2}$ quantum antiferromagnetic $XY$ model on finite
triangular lattices with $N$ sites
in both longitudinal and transverse magnetic fields.
We calculate physical quantities in the ground state
using a diagonalization for spins $N \leq 27$,
and those at finite temperatures
using a quantum transfer Monte Carlo method for $N \leq 24$.
In the longitudinal magnetic field,
the long-range chiral order parameter seems to have a finite,
nonzero value at low temperatures
suggesting the occurrence of a classical umbrella-type phase.
In the transverse magnetic field,
the 1/3-plateau of the magnetization curve appears
even at low temperatures,
in contrast with the classical model.
The magnetic field dependences of the order parameters suggest that
the chiral-ordered, the ferrimagnetic,
and the spin flop phases appear successively
as the magnetic field is increased.
The transition temperatures are estimated
from the peak position of the specific heat,
and the phase diagrams are predicted
in both longitudinal and transverse magnetic fields.
\end{abstract}

\narrowtext

\section{INTRODUCTION}

The antiferromagnetic $XY$ model on the triangular lattice
has attracted much interest.
The Hamiltonian is
\begin{eqnarray}
{\cal H}=2J\sum_{\langle i,j \rangle}(S_{i}^{x}S_{j}^{x}+S_{i}^{y}S_{j}^{y})
- H^\alpha \sum_{i}S_{i}^{\alpha}, \label{hamiltonian}
\end{eqnarray}
where $J$ $(> 0 )$ is the exchange integral,
$H^\alpha$ is an external magnetic field along the $\alpha$ direction,
and $\langle i,j \rangle$ takes all the nearest neighbor pairs.
In the classical model,
Miyashita and Shiba\cite{miyashita84}
found using the Monte Carlo method
that a long-range
chiral-ordered phase exists at low temperatures,
although the long-range order of the spins disappears
at finite temperatures.
They also suggested that the transition belongs to the universality
class of the Ising model.
The model in magnetic fields was also studied.
In a longitudinal field,
all the spins lean to the z direction
but the chiral-ordered phase still exists
(Fig.~\ref{candidate}, the umbrella-type phase).
On the other hand,
in a transverse magnetic field,
Lee et al.\ showed that the chiral-ordered,
the ferrimagnetic, and the spin flop phases appear successively
as the magnetic field increases
(in Figs.~\ref{candidates}(a)-(c)).~\cite{lee86}
In the ferrimagnetic phase,
the magnetization has a value of 1/3 of the saturated value,
and the magnetization curve exhibits a plateau,
which is often called as a 1/3-plateau.
However, the ferrimagnetic region of the field shrinks
as the temperature cools down,
and vanishes at $T = 0$.

In the $S = 1/2$ quantum model, it has been discussed
whether the classical ordered phases are broken down
by the quantum fluctuation or not.~\cite{%
fujiki87,nishimori88,leung93,suzuki96,matsubara88,fujiki91,momoi92a,momoi92b}
The ground state properties of the model on finite lattices have 
been studied 
using a diagonalization method.~\cite{fujiki87,nishimori88,leung93,suzuki96}
The largest lattice treated up to now is that with $N = 36$,~\cite{leung93}
and the results suggest that the chiral order parameter remains finite for
$N \rightarrow \infty$.
The properties at finite temperatures were also studied by several authors 
using a quantum transfer Monte Carlo (QTMC),~\cite{matsubara88,imada86}
a high temperature series expansion,~\cite{fujiki91}
and a super-effective-field theory.~\cite{momoi92a}
Recently,
the authors carefully studied an anisotropic Heisenberg model
using the QTMC method and showed that,
when $XY$ anisotropy is large,
the peak-height of the specific heat increases with $N$.~\cite{suzuki95}
Thus they predicted that the quantum fluctuation does not destroy
the classical chiral-ordered phase in finite temperatures.
In presence of magnetic fields, on the contrary,
it was speculated that the quantum fluctuation brings
a different ground state from the classical one.
Chubukov and Golosov studied the ground state of the $XY$ model
in the transverse field
using the spin wave theory,
and suggested that the 1/3-plateau appears at $T = 0$.~\cite{chubukov91}
However,
the spin wave theory does not always give a correct result
in the frustrated model.~\cite{chandra88}
Moreover, when $S = 1/2$,
the critical field at which the plateau appears becomes negative
because of the $1/S$-expansion.~\cite{chubukov91}

In this paper, we study the $S = 1/2$ quantum $XY$ model in magnetic fields.
We consider the magnetic structures both at $T = 0$ and at $T \neq 0$.
We calculate quantities in the ground state of finite lattices
with $N \leq 27$ by the diagonalization method. 
On the other hand, those in finite temperatures are calculated
on the lattice with $N \leq24$ using the QTMC method.
From the result, we suggest the following things.
(1) In the longitudinal field,
the umbrella-type phase occurs up to $H^z = 3.0$.
(2) In the transverse field,
the 1/3-plateau really appears even in the ground state.
The chiral-ordered, the ferrimagnetic, and the spin flop phases occur
as the magnetic field is increased.
The phase diagrams in both longitudinal and transverse fields
are predicted.

We consider the models in the longitudinal and the transverse fields
in Sec.~II and Sec.~III, respectively.
Summary is given in Sec.~IV.
We explain the QTMC method in Appendix A.

\section{LONGITUDINAL MAGNETIC FIELD}

In this section, we study the model in a longitudinal field $H^z$.
First, we consider the magnetization and the chiral order parameter
in the ground state.
The total magnetization along the external field $H^z$ is given by
\begin{eqnarray}
M^z = (1/N) \sum_i S^z_i.
\end{eqnarray}
The z component of the chirality of the upright triangle at
\mbox{\bf R} is defined as
\begin{eqnarray}
\chi^{z}({\bf R})&=&\frac{1}{\sqrt{3}}(S_{i}^{x}S_{j}^{y}
-S_{i}^{y}S_{j}^{x}+S_{j}^{x}S_{k}^{y}-S_{j}^{y}S_{k}^{x} \nonumber \\
&+&S_{k}^{x}S_{i}^{y}
-S_{k}^{y}S_{i}^{x}), \label{chiz}
\end{eqnarray}
where the relation of $i \rightarrow j \rightarrow k$
is taken counterclockwise, as is shown in Fig.~\ref{candidates}(a).
The eigenvalues of $\chi^z({\bf R})$ are $\pm1/2$ and 0.
The long-range chiral order parameter is defined as
\begin{eqnarray}
\chi^2 = \frac{1}{NS(NS + 1)}[\sum_{{\bf R} \in \Delta}%
\chi^{z}({\bf R})]^2, \label{chiral}
\end{eqnarray}
where {\bf R} runs over all the upright triangles on the lattice.
In Figs.~\ref{mzvshzg} and \ref{chivshzg},
we show $M^z$ and $\sqrt{\chi^2}$,
respectively.
Since their size dependences are almost negligible,
these values will be similar to those for $N \rightarrow \infty$.
The magnetization increases almost linearly
with $H^z$, and no anomaly is observed
until the saturated field $H^z_{max} = 3.0$.
The chiral order parameter decreases monotonically
with increasing $H^z$, and vanishes at $H^z_{max} = 3.0$.
These results suggest that the chiral-ordered phase
changes into an umbrella-type one,
and it survives up to $H^z_{max} = 3.0$.
If it is true,
the chiral order parameter at $H^z \neq 0$
will be approximately given as
$\sqrt{\tilde\chi^2} = \sqrt{\chi^2}_{H^z = 0}\cos^2\theta %
=  \sqrt{\chi^2}_{H^z = 0}(1-(M^z)^2)$,
where $\theta$ is the angle between the spin and $XY$-plane.
We calculate $\sqrt{\tilde\chi^2}$ and show it
in Fig.~\ref{chifrommz} together with $\sqrt{\chi^2}$.
The two results are in good agreement with each other.
Thus we may conclude that in the ground state
the umbrella-type phase occurs at $H^z \neq 0$,
and it survives up to $H^z_{max} = 3.0$.

Next we consider $\langle M^z \rangle$
and $\sqrt{\langle\chi^2\rangle}$ for a finite temperature,
where $\langle \cdots \rangle$ denotes the thermal average.
All the results presented below have been obtained by the QTMC method
(see Appnexdix A).
The number of the states $M$ used in the QTMC calculation are as follows:
$M = 50$ for $N = 9-18$, $M = 10$ for $N = 21$, and $M = 2$ for $N = 24$.
For every lattice except for $N = 24$,
the set of $M$ states is divided into five subsets,
and quantities of interest are calculated in every subset.
Error bars presented in figures given below only mean
deviations of the values obtained in different subsets.
In the QTMC calculation,
since the operators ${\cal H}_0 = 2J\sum (S^x_iS^x_j + S^y_iS^y_j)$ and 
${\cal H}_1 = -H^z\sum S^z_i$ are commutable,
$\exp [-({\cal H}_0 + {\cal H}_1)/T]$ may be decomposed into
$\exp [-{\cal H}_0/T]\exp[-{\cal H}_1/T]$.
Therefore,
we can calculate the quantities for a finite temperature
combining the $T$ transfer and the $H^z$ transfer.
That is, those quantities are transferred from a high temperature,
$T_0$, to a temperature of interest,
$T_{fin}$,
by operating $\exp [-{\cal H}_0/T_0]$.
Then, the field is increased by the operator $\exp [-{\cal H}_1/T_{fin}]$.
In Figs.~\ref{mzvshz} and \ref{chivshz},
we show $\langle M^z \rangle$ and $\sqrt{\langle\chi^2\rangle}$
at $T = 0.2$ for different $N$,
respectively.
In low fields,
the results are very similar to those of the ground state and
their size dependences are small.
As $H^z$ is increased beyond $H^z = 2.5$,
their size dependences become considerable.
Therefore, we suggest that
at $T = 0.2$ the umbrella-type phase survives at least $H^z \lesssim 2.5$.

We also consider the temperature dependence of the specific heat $C$,
$\langle M^z \rangle$, and $\sqrt{\langle\chi^2\rangle}$ at fixed fields.
The specific heat is calculated from
\begin{eqnarray}
C= \frac{1}{T^{2}} %
( \langle {\cal H}^2 \rangle - \langle {\cal H} \rangle ^{2} ).
\end{eqnarray}
We found considerable differences
in those quantities between for odd $N$ and for even $N$,
but they are reduced with increasing $N$.
In Fig.~\ref{cwithhz}(a), we plot $C$ as a function of $T$ at $H^z = 1.0$.
As $N$ increases, the peak of $C$ becomes higher
and sharper at $T \sim 0.4$.
A similar size dependence can be seen up to $H^z \sim 2.0$.
Results at $H^z = 2.0$ are also shown in Fig.~\ref{cwithhz}(b),
in which the increase of $C$ is seen at low temperatures.
In Figs.~\ref{mzvst} and \ref{chivst},
we plot $\langle M^z \rangle$
and $\sqrt{\langle\chi^2\rangle}$ as functions of $T$.
As the temperature is lowered,
$\langle M^z \rangle$ increases slowly and saturates at $T \sim 0.4$.
$\sqrt{\langle\chi^2\rangle}$ increases
and its size dependence becomes smaller,
which indicates that $\sqrt{\langle\chi^2\rangle}$
remains finite in the thermodynamic limit.
Therefore all the results suggest
that the phase transition occurs at least $H^z \leq 2.0$.

The phase diagram of the model predicted
from the above results is shown in Fig.~\ref{phasehz}.
The transition line is estimated from the peak position of $C$.
The chiral-ordered phase
with the longitudinal magnetization (the umbrella-type phase)
survives in the longitudinal magnetic field $H^z$.

\section{TRANSVERSE  MAGNETIC FIELD}

Next we study the model in the transverse field $H^x$.
Since the total magnetization $M^z$ is not good quantum number,
we must treat all the $2^N$ states in the diagonalization method.
The QTMC method prepares all the Ising states,
and it does not suffer from this difficulty.
We calculate the magnetization,
the order parameter defined below, and the specific heat.
The number of the states $M$ are as follows:
$M = 50$ for $N \leq 18$, $M = 10$ for $N = 21$,
and $M = 2$ for $N = 24$.

We calculate the total magnetization given by
\begin{eqnarray}
M^x = \sum_i S^x_i / N.
\end{eqnarray}
This quantity becomes 1/6 if the ferrimagnetic phase appears
and 1/2 if all the spins align along the x direction.
In Figs.~\ref{mxvshx}(a) and (b),
we show the magnetization
as functions of $H^x$ at $T = 0.133$ and 0.333, respectively.
The data for different $N$ lie almost on the same line,
and the temperature dependence is negligibly small.
Note that we also calculate $\langle M^x \rangle$
in the ground state for $N = 12$,
and find the result is quite similar to that at $T = 0.133$.
There is the 1/3-plateau
in the region of $2.0 \lesssim H^x \lesssim 4.0$.
We estimate two edges of the plateau $H_{c1}$ and $H_{c2}$
at various temperatures,
and plot them in the phase diagram shown below.
These edges depend little on the temperature.
Note that even at $T = 0.133$,
$\langle M^x \rangle$ slightly increases
with $H^x$ between $H_{c1}$ and $H_{c2}$,
in contrast with
that of the Heisenberg model,~\cite{chubukov91,nishimori86}
because $\langle M^x \rangle$
is not good quantum number in the $XY$ model.

Now we consider the spin structure.
In Fig.~\ref{candidates}, we show three possible spin structures
(a), (b), and (c) which appear in the classical model at low,
intermediate, and high fields, respectively.
These structures are called as the chiral-ordered,
the ferrimagnetic, and the spin flop phases, respectively.
In order to see which structure is realized with $H^x$,
we calculate the chiral order parameter
and x and y components of the sublattice order parameters
which are defined as
\begin{eqnarray}
{\cal M}^\alpha = [\frac{1}{2}\langle (&M^\alpha_A& - M^\alpha_B)^2%
     + (M^\alpha_B - M^\alpha_C)^2 \nonumber \\%
     + (&M^\alpha_C& - M^\alpha_A)^2\rangle ]^{1/2}
\end{eqnarray}
where $M^\alpha_\zeta(\equiv \sum_{i \in \zeta}S_i^\alpha/(N/3))$ means
the $\alpha$ component of the magnetization of the $\zeta$ sublattice.
In Table.~\ref{relation} we show the relation
between the order parameters and the spin structures.
In Figs.~\ref{chivshx}-\ref{my2vshx},
we plot $\sqrt{\langle\chi^2\rangle}$,
${\cal M}^x$, and ${\cal M}^y$, respectively,
as functions of $H^x$ at $T = 0.133$.
When $H^x < H_{c1}$,
their size dependences are not so large.
Especially, for $H^x \leq 1.0$,
$\sqrt{\langle\chi^2\rangle}$ depends very little on $H^x$.
As $H^x$ reaches $H_{c1}$,
$\sqrt{\langle\chi^2\rangle}$ and ${\cal M}^y$ decrease markedly
and their size dependences become larger.
On the contrary, ${\cal M}^x$ increases and its size dependence becomes small.
When $H_{c1} < H^x < H_{c2}$,
$\sqrt{\langle\chi^2\rangle}$ and ${\cal M}^y$
remarkably depend on the system sizes.
On the other hand, the size dependence of ${\cal M}^x$ is negligible.
When $H^x > H_{c2}$,
${\cal M}^x$ decreases and its size dependence is considerable.
${\cal M}^y$ increases again and its size dependence is smaller,
but $\sqrt{\langle\chi^2\rangle}$ still decreases.
We tentatively assume that the data fit on the $1/\sqrt{N}$-function,
and estimated the values for $N \rightarrow \infty$
which are also shown in the figures.
When $H^x < H_{c1}$, $\sqrt{\langle\chi^2\rangle}$,
${\cal M}^x$, and ${\cal M}^y$ have a finite, nonzero value.
When $H_{c1} < H^x < H_{c2}$, $\sqrt{\langle\chi^2\rangle}$, and ${\cal M}^y$
seem to vanish for $N \rightarrow \infty$,
whereas ${\cal M}^x$ remains finite.
When $H_{c2} < H^x$, ${\cal M}^x$ and ${\cal M}^y$ have a finite,
nonzero values, but $\sqrt{\langle\chi^2\rangle}$ disappears.
From these results,
we suggest that the structures (a), (b), and (c)
in Fig.~\ref{candidates} occur
when $H^x < H_{c1}$,
$H_{c1} < H^x < H_{c2}$, and $H_{c2} < H^x$, respectively.
Of course, these picture of the spin structure are
compatible with the magnetization curve
which are shown in Figs.~\ref{mxvshx} 

In Figs.~\ref{cwithhx}(a)-(d), we present the specific heat $C$
at $H^x = 1.0$, 3.0, 5.0, and 7.0.
For $H^x \leq 5.0$, the peak height becomes higher with increasing $N$
suggesting the occurrence of the phase transition.
At $H^x = 7.0$, the peak height does not increase with $N$,
but $C$ does at low temperatures.
Therefore we expect
that the phase transition also takes place at $H^x = 7.0$.
The transition temperatures are estimated
from the size dependence of the peak-temperature $T_m$.~\cite{suzuki95}

In Fig.~\ref{phasehx}, we show the phase diagram of the model.
This phase diagram is analogous to that of the classical model.
However, it should be emphasized that
the 1/3-plateau widely survives at the low temperatures.

\section{SUMMARY}

In this paper,
we have studied the $S = 1/2$ quantum antiferromagnetic $XY$ model
on finite triangular lattice
in both longitudinal and transverse magnetic fields
using the diagonalization and the QTMC methods.
We have calculated the magnetization,
the specific heat, and the order parameters
for different sizes of the lattices,
and examined their size dependences to see
whether some long-range ordered phase occurs or not.
Our results are summarized as follows.

In the longitudinal magnetic field,
the long-range chiral order parameter
seems to have a finite, nonzero value at low temperatures.
The peak height of the specific heat increases with $N$
suggesting the occurrence of the phase transition.
From the results, we suggest
that the umbrella-type phase survives at low temperatures,
and predict the phase diagram.

In the transverse magnetic field,
the 1/3-plateau is seen
in the magnetization curve at low temperatures,
in contrast with the classical model.
The behavior of the order parameters suggests
that the chiral-ordered, the ferrimagnetic,
and the spin flop phases appear successively
as the field is increased.
The peak height of the specific heat increases with $N$
suggesting the occurrence of the long-range order.
Thus we suggest that the three phases occur
in the transverse magnetic field
at finite temperatures as in the classical model.
The quantum fluctuation destroys
the order of the y component of the spin,
and the ferrimagnetic phase appears at low temperatures.

\acknowledgments
We would like to thank Dr.\ T. Nakamura for valuable discussions.
The diagonalization programs are based
on the subroutine package ``TITPACK Ver.\ 2''
coded and supported by Professor H. Nishimori,
and ``KOBEPACK/S Ver.\ 1.1'' by Professor T. Tonegawa,
Professor M. Kaburagi and Dr.\ T. Nishino.
The computation in this work has been done using the
facilities of the Supercomputer Center, Institute for Solid
State Physics, University of Tokyo.
This work was supported by Grant-in-Aid for Scientific Research
from the Ministry of Education,
Science and Culture.

\appendix

\section{The QTMC method}

In QTMC method,~~\cite{imada86,suzuki95} 
we calculate the following quantity:
\begin{equation}
\widetilde{\langle A \rangle} =\sum_{k}^{M}\langle \psi_{k}|A\exp(- \beta H)|
\psi_{k}\rangle /\sum_{k}^{M}\langle \psi_{k}|\exp(- \beta H)|\psi_{k}\rangle
\label{approx}
\end{equation}
where $A$ is some physical operator and
the sum runs over $M$ states each of which is given by
\begin{equation}
|\psi_k\rangle =\sqrt{\frac{6}{M}}\sum_{i}^{2^{N}}C_{ik}|i\rangle  \label{psi},
\end{equation}
here $C_{ik}$ is a random number of $-1 \leq C_{ik} \leq 1$. We can readily
show that $\widetilde{\langle A \rangle} \rightarrow \langle A \rangle$ for
$M \rightarrow \infty$, because $(6/M) \displaystyle\sum_{k}^{M} C_{ik}C_{jk} =
\delta_{ij}+O(1/\sqrt{M})$.~\cite{suzuki95}
A great advantage
of using the formula~(\ref{approx}) is that we can obtain an approximate value
of the average $\langle A \rangle$ by summing up only $M$ terms instead of all
the $2^{N}$ terms of the Ising states. Using this formula, we can treat systems
much larger than those treatable by the rigorous formula. Of course,
statistical errors of $O(1/ \sqrt{M})$  arise, but we may reduce them as $M$ is
increased.

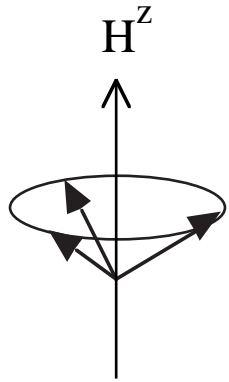
\begin{figure}
\caption{A spin structure of the classical model
in the longitudinal magnetic field,
which is called by the umbrella-type phase.}
\label{candidate}
\end{figure}
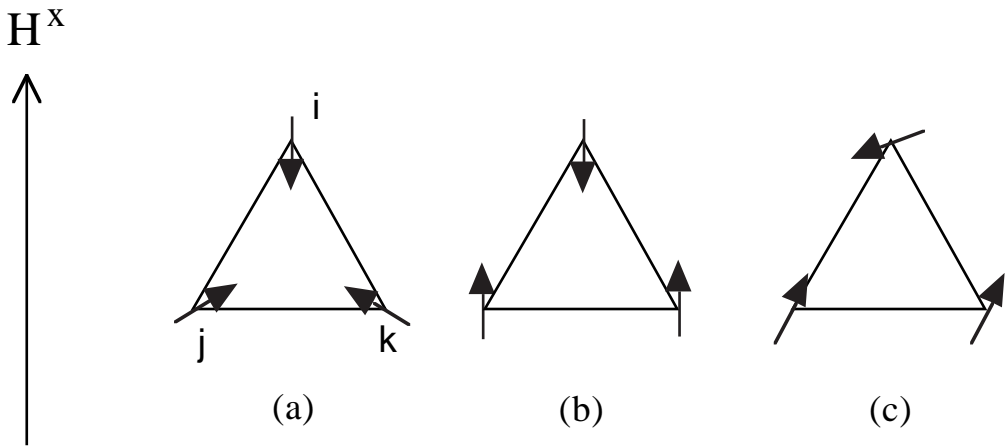
\begin{figure}
\caption{Three spin structures of the classical model
in the transverse magnetic field.
(a)~The chiral-ordered phase,
(b)~the ferrimagnetic phase, and (c)~the spin flop phase.}
\label{candidates}
\end{figure}

\begin{figure}
\caption{The field dependence of the magnetization $M^z$ at $T = 0$.}
\label{mzvshzg}
\end{figure}
\begin{figure}
\caption{The field dependence of the chiral order parameter
$\protect\sqrt{\chi^2}$ at $T = 0$.}
\label{chivshzg}
\end{figure}
\begin{figure}
\caption{$\protect\sqrt{\tilde\chi^2}$
and $\protect\sqrt{\chi^2}$ at $T = 0$.}
\label{chifrommz}
\end{figure}
\begin{figure}
\caption{The field dependence of $\langle M^z\rangle$ at $T = 0.2$.}
\label{mzvshz}
\end{figure}
\begin{figure}
\caption{The field dependence of
$\protect\sqrt{\langle\chi^2\rangle}$ at $T = 0.2$.}
\label{chivshz}
\end{figure}
\begin{figure}
\caption{The temperature dependence of the specific heat $C$ for different
$H^z$: (a)~$H^z = 1.0$ and (b)~$H^z = 2.0$.}
\label{cwithhz}
\end{figure}
\begin{figure}
\caption{The temperature dependence of $\langle M^z\rangle$.}
\label{mzvst}
\end{figure}
\begin{figure}
\caption{The temperature dependence of
$\protect\sqrt{\langle\chi^2\rangle}$ for different $H^z$: 
(a)~$H^z = 1.0$ and (b)~$H^z = 2.0$.}
\label{chivst}
\end{figure}
\begin{figure}
\caption{Phase diagram in longitudinal fields.
The symbol $\times$ is estimated
from the peak position of the specific heat.
The line is a guide to the eye.}
\label{phasehz}
\end{figure}

\begin{figure}
\caption{The field dependence of the magnetization
$\langle M^x\rangle$ for different $T$:
(a)~$T = 0.133$ and (b)~$T = 0.333$.}
\label{mxvshx}
\end{figure}
\begin{figure}
\caption{The chiral order parameter
$\protect\sqrt{\langle\chi^2\rangle}$ vs.\ $H^x$.
The symbol $\bullet$ denotes the extrapolated value.}
\label{chivshx}
\end{figure}
\begin{figure}
\caption{The order parameter ${\cal M}^x$ vs.\ $H^x$.
The symbol $\bullet$ denotes the extrapolated value.}
\label{mx2vshx}
\end{figure}
\begin{figure}
\caption{The order parameter ${\cal M}^y$ vs.\ $H^x$.
The symbol $\bullet$ denotes the extrapolated value.}
\label{my2vshx}
\end{figure}
\begin{figure}
\caption{The temperature dependence of the specific heat $C$ for different
$H^x$: (a)~$H^x = 1.0$, (b)~$H^x = 3.0$,
(c)~$H^x = 5.0$, and (d)~$H^x = 7.0$.}
\label{cwithhx}
\end{figure}
\begin{figure}
\caption{Phase diagram in transverse fields.
The symbols $\times$ and $\triangle$
are estimated from the edges of the plateau
of the magnetization curve and
the peak height of the specific heat, respectively.
The lines are guides to the eye.}
\label{phasehx}
\end{figure}

\begin{table}
\begin{tabular}{lccc}
                & Structure (a) & Structure (b) & Structure (c) \\
                \hline
$\sqrt{\langle\chi^2\rangle}$ & $\neq 0$ & $ = 0$ & $ = 0 $ \\
${\cal M}^x$    & $\neq 0$ & $\neq 0$ & $\neq 0$ \\
${\cal M}^y$    & $\neq 0$ & $ = 0$   & $\neq 0$
\end{tabular}
\caption{The relation between
the order parameters and the classical spin structures.}
\label{relation}
\end{table}

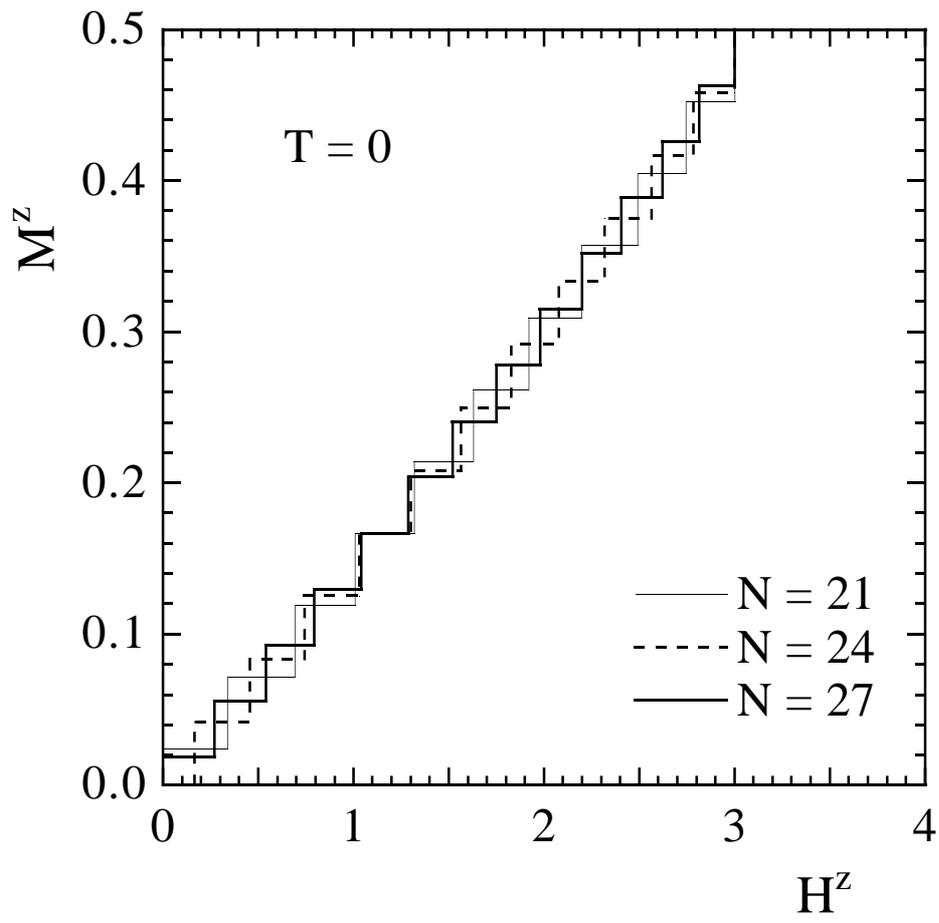
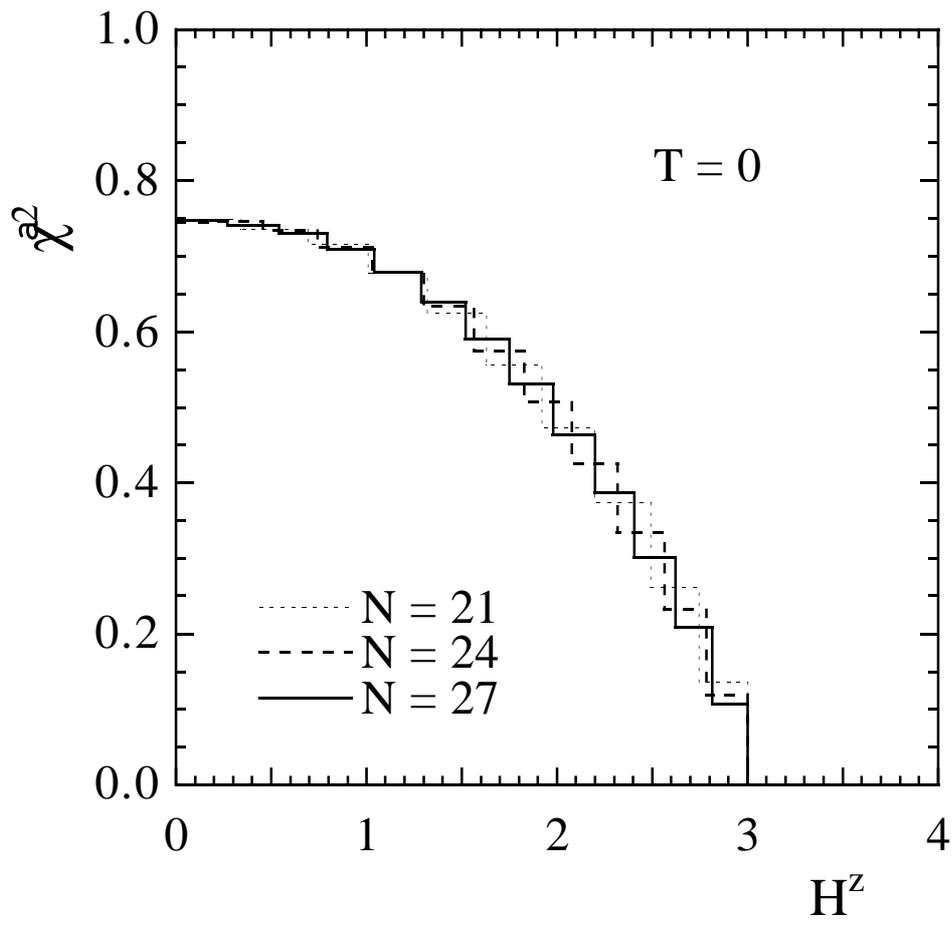
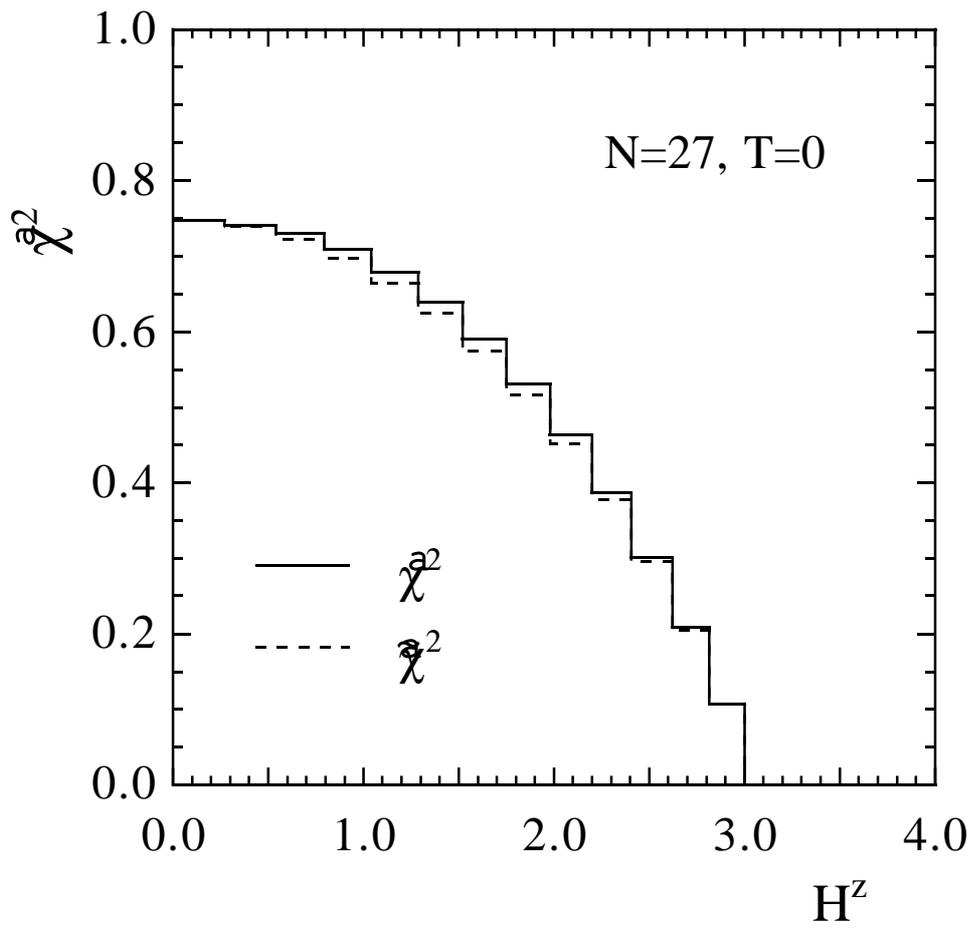
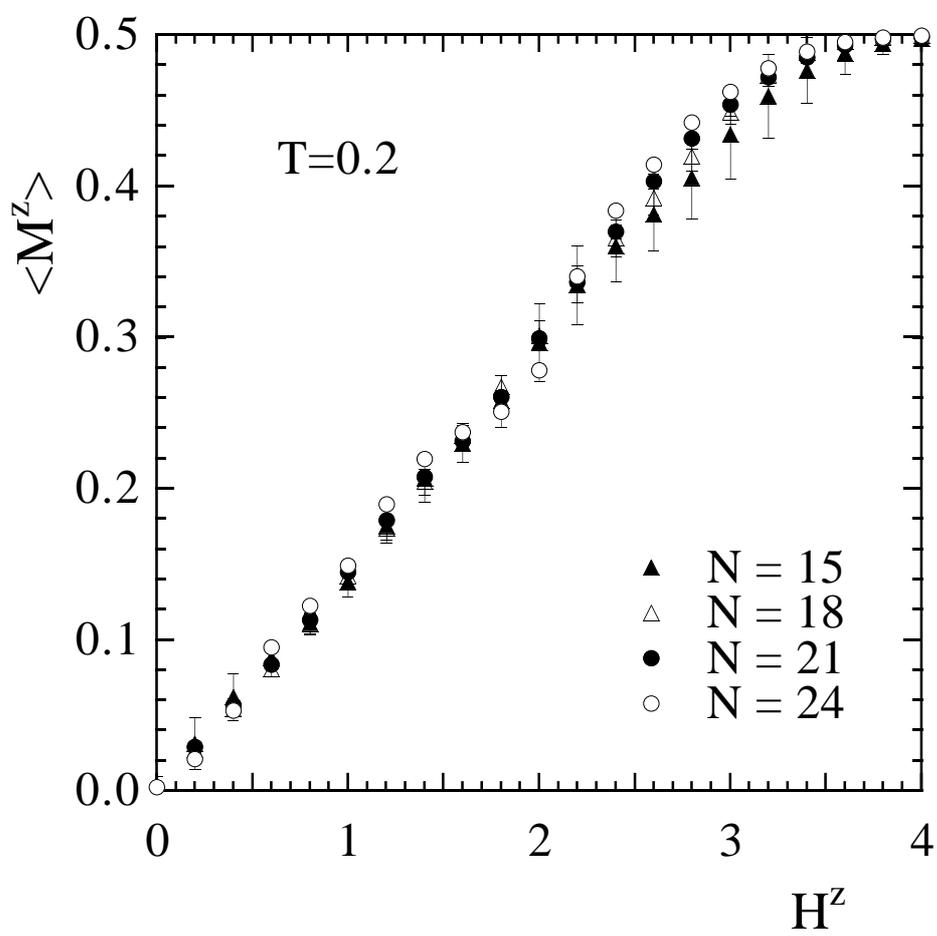
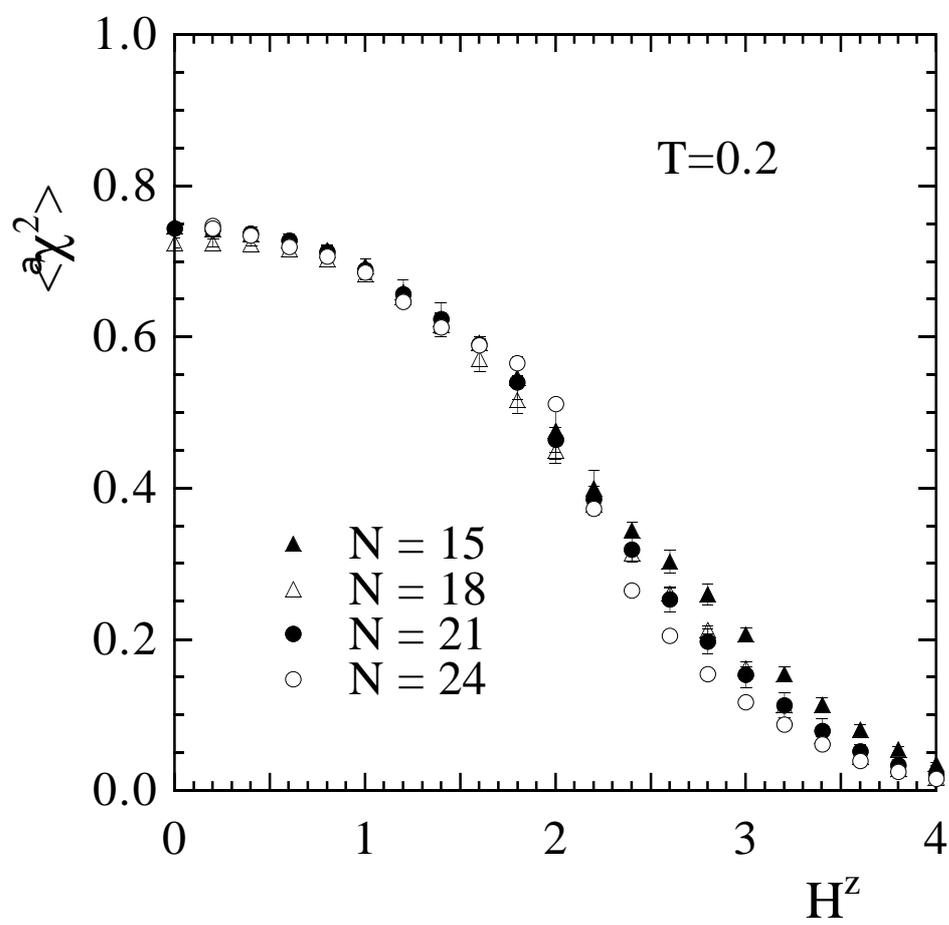
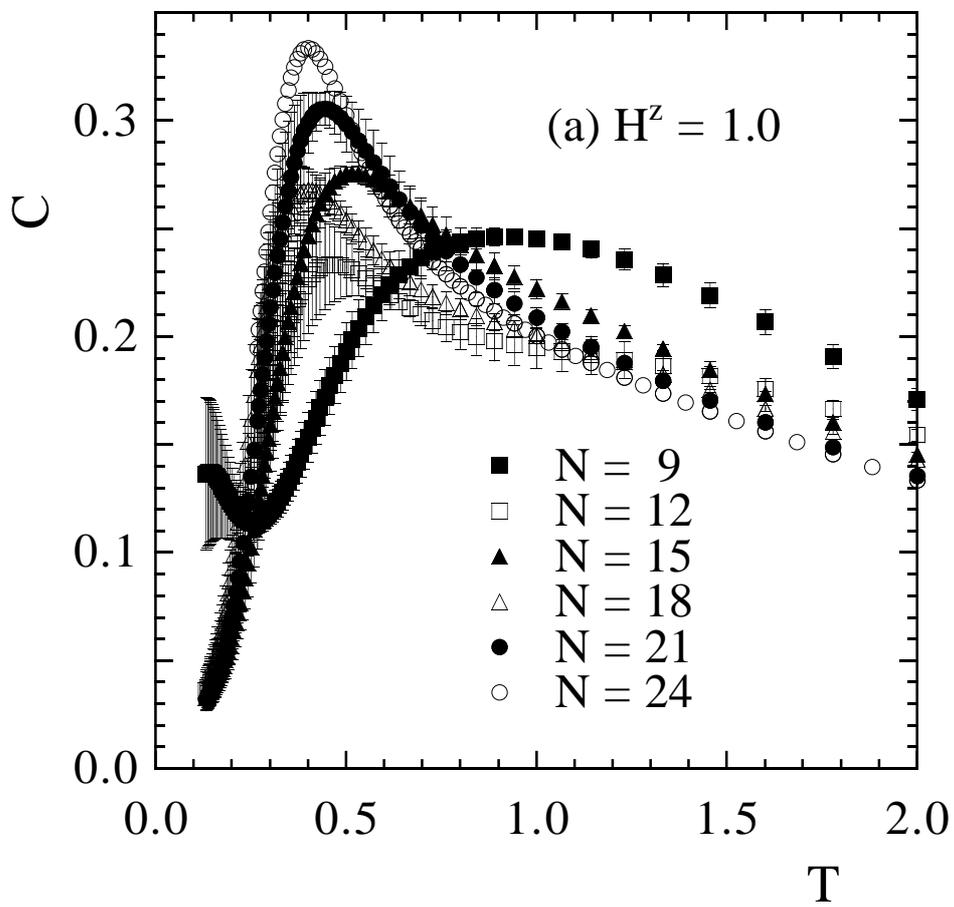
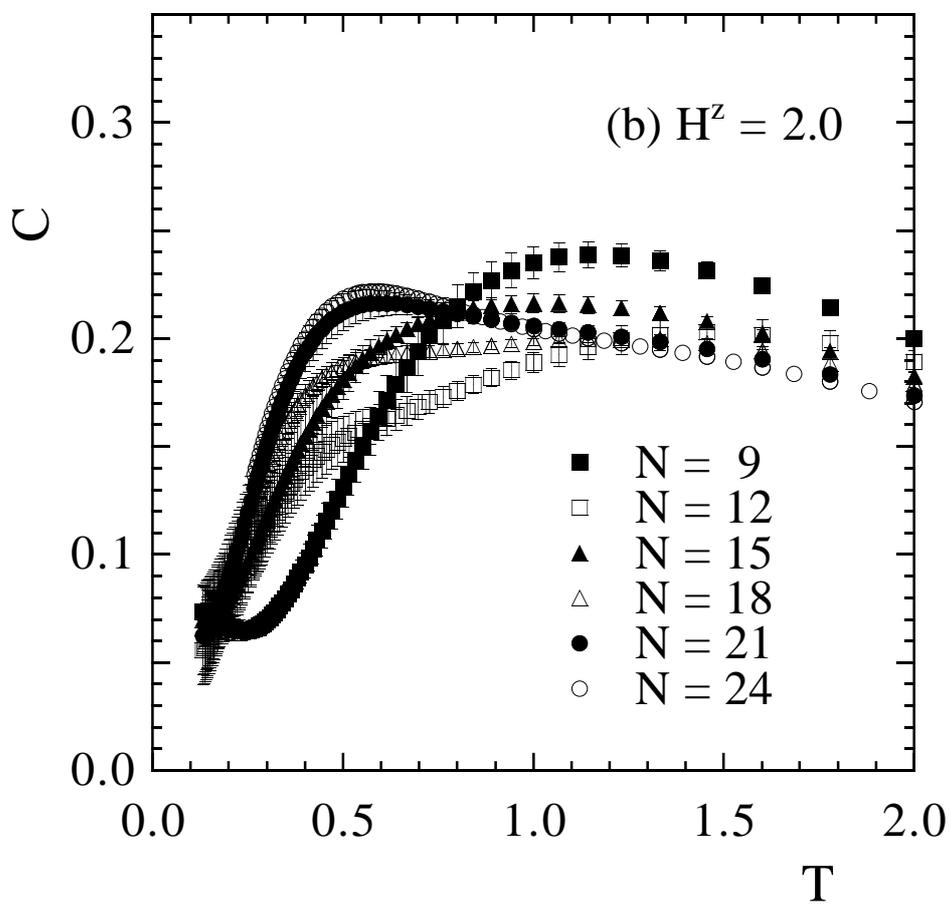
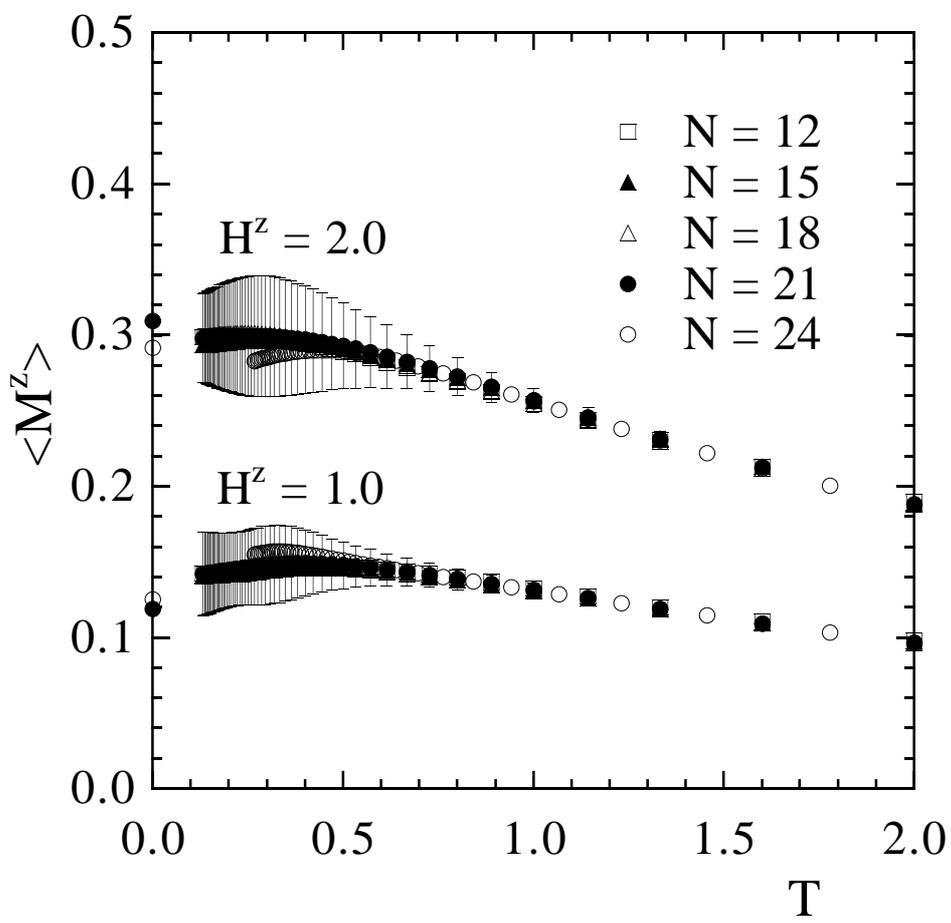
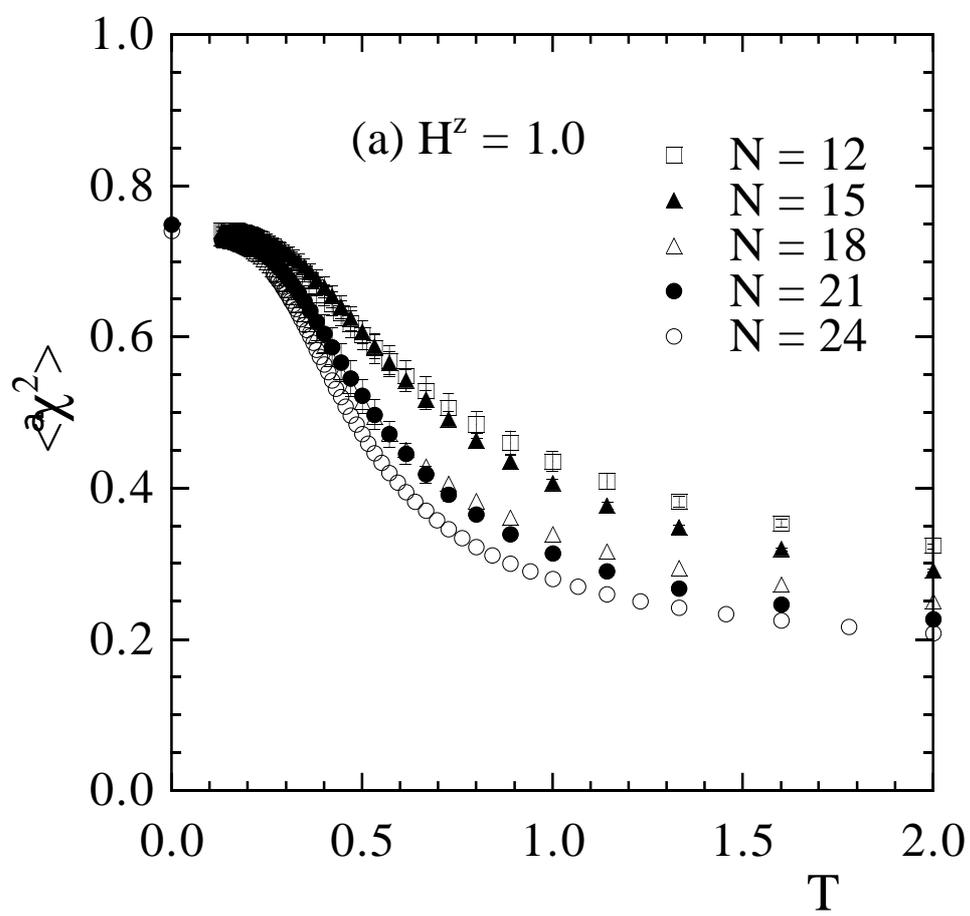
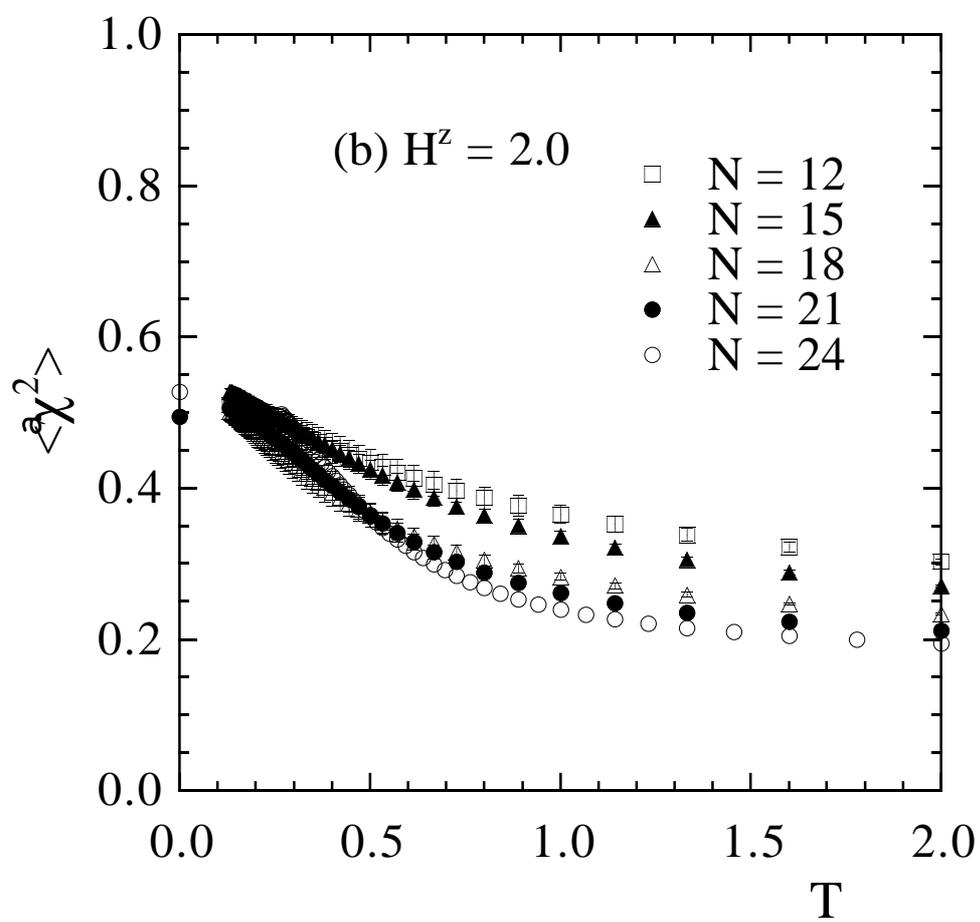
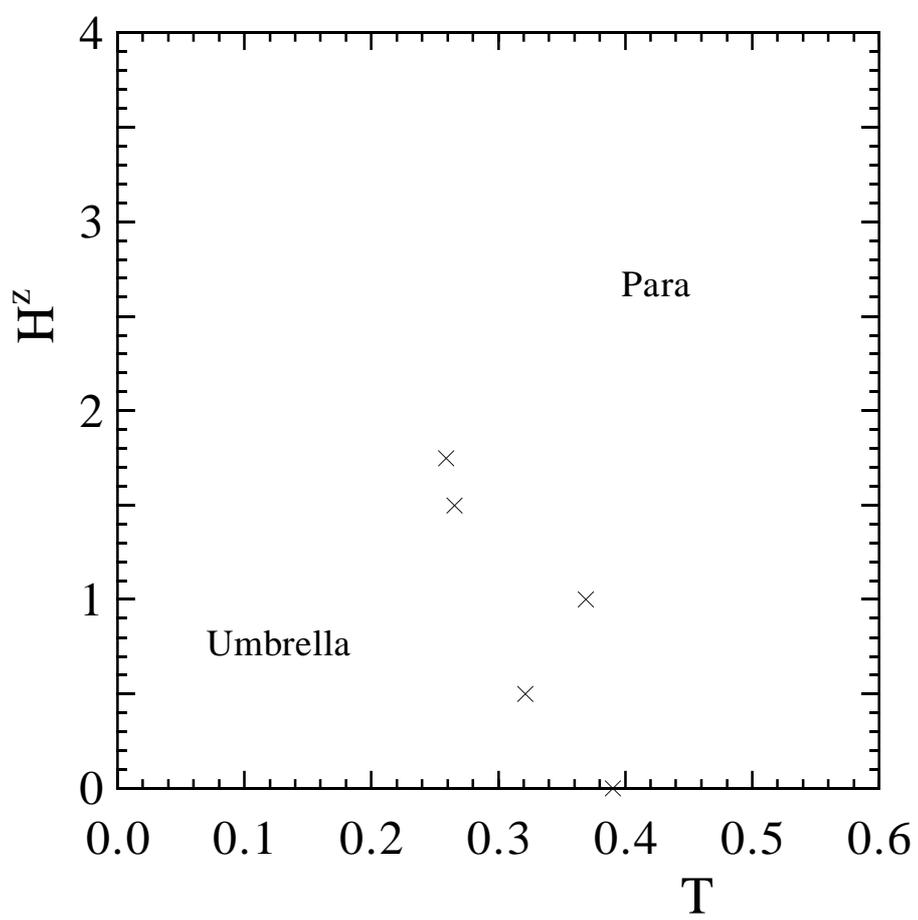
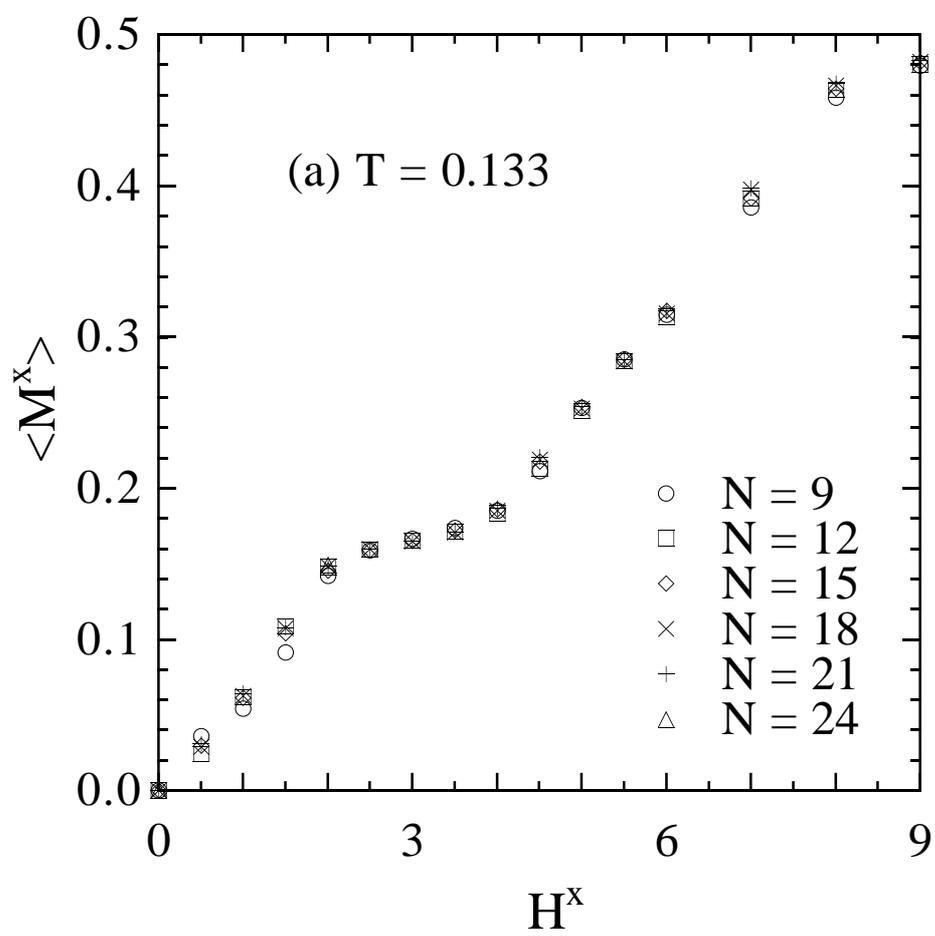
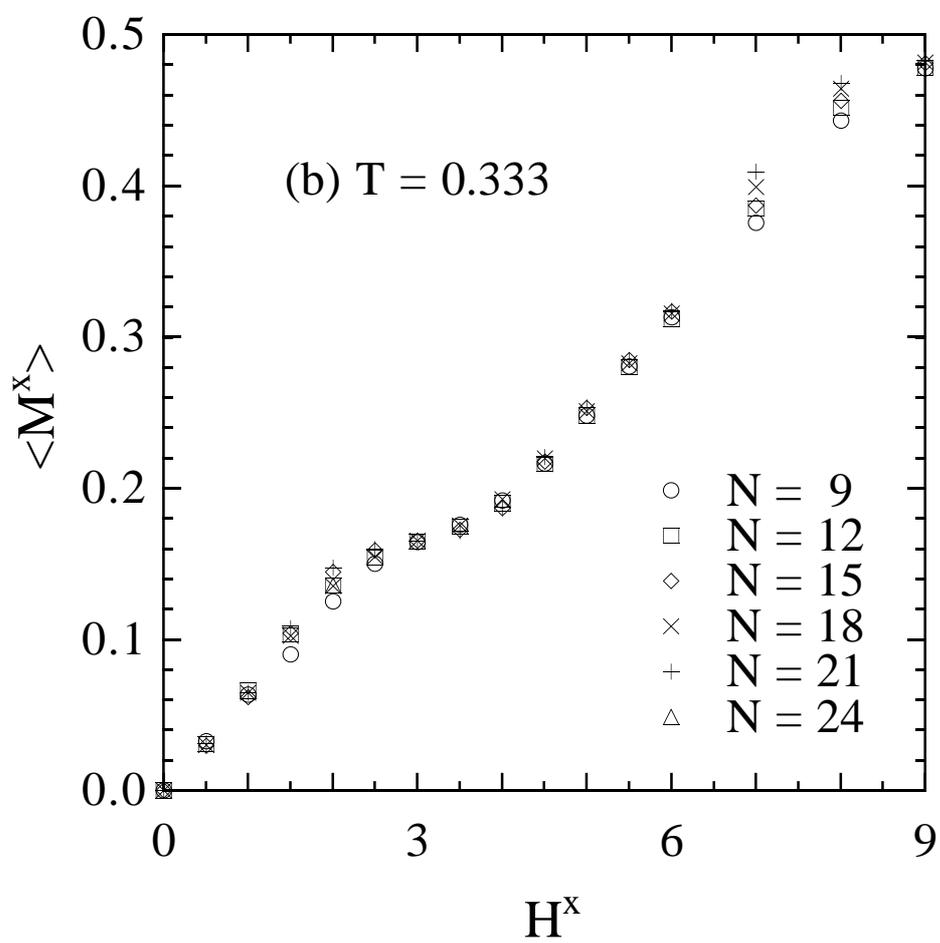
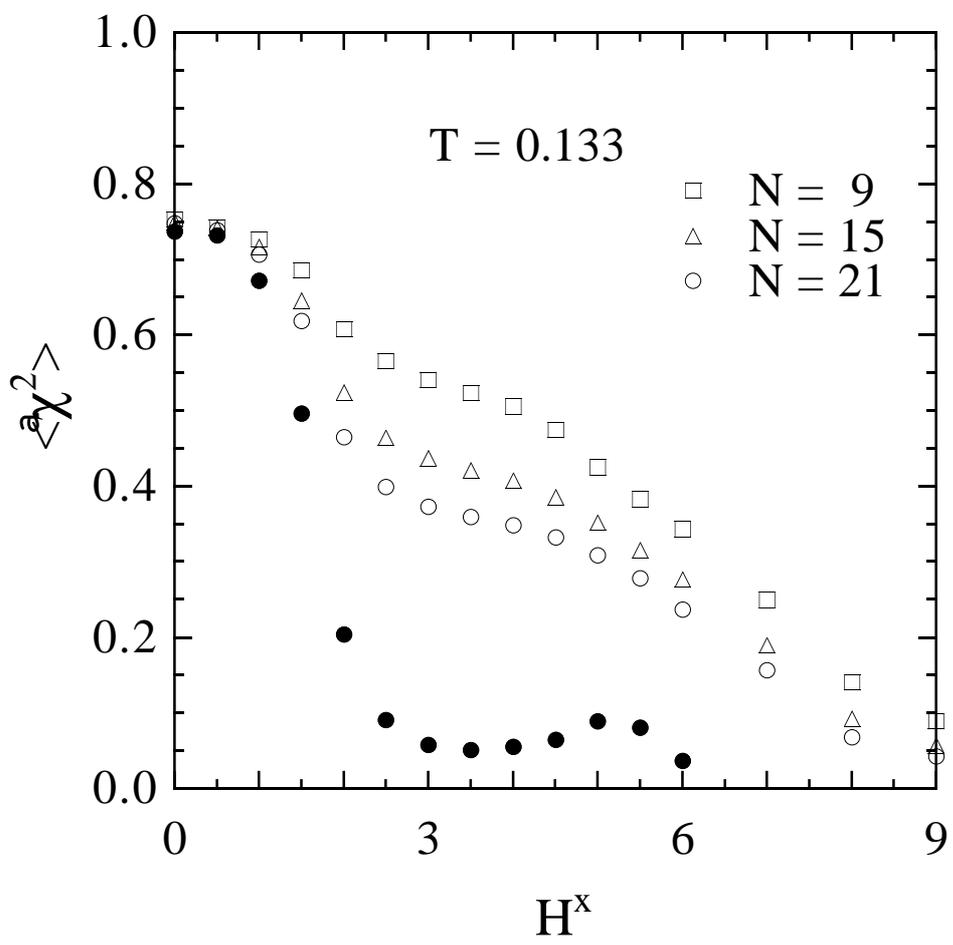
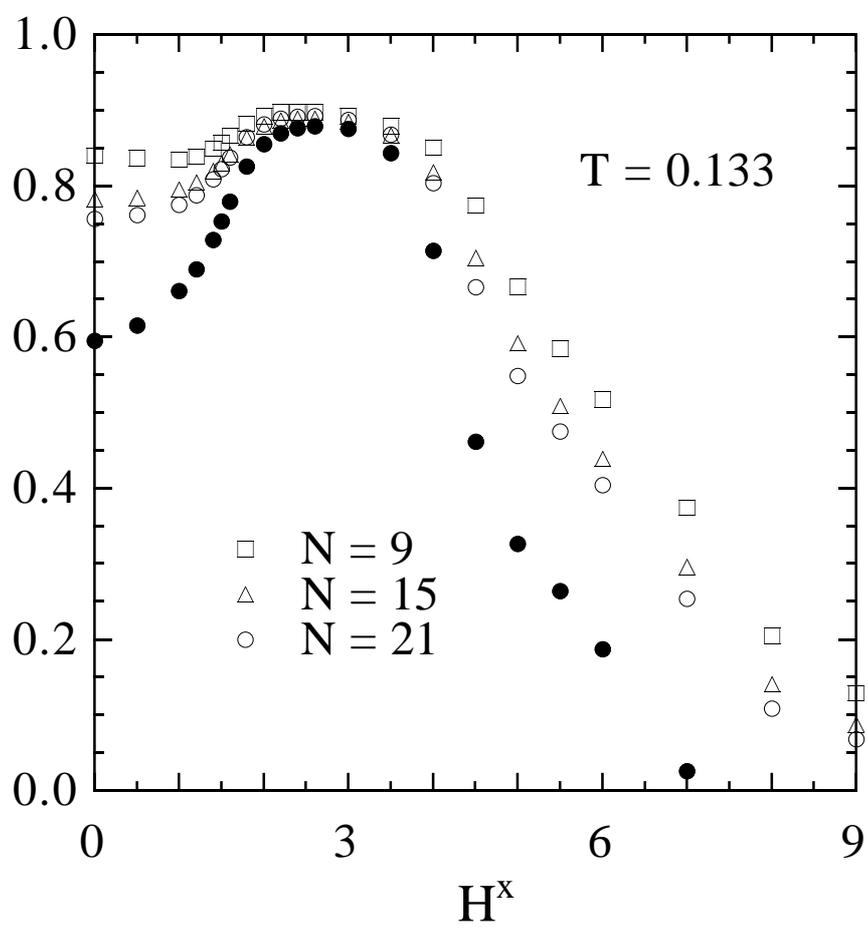
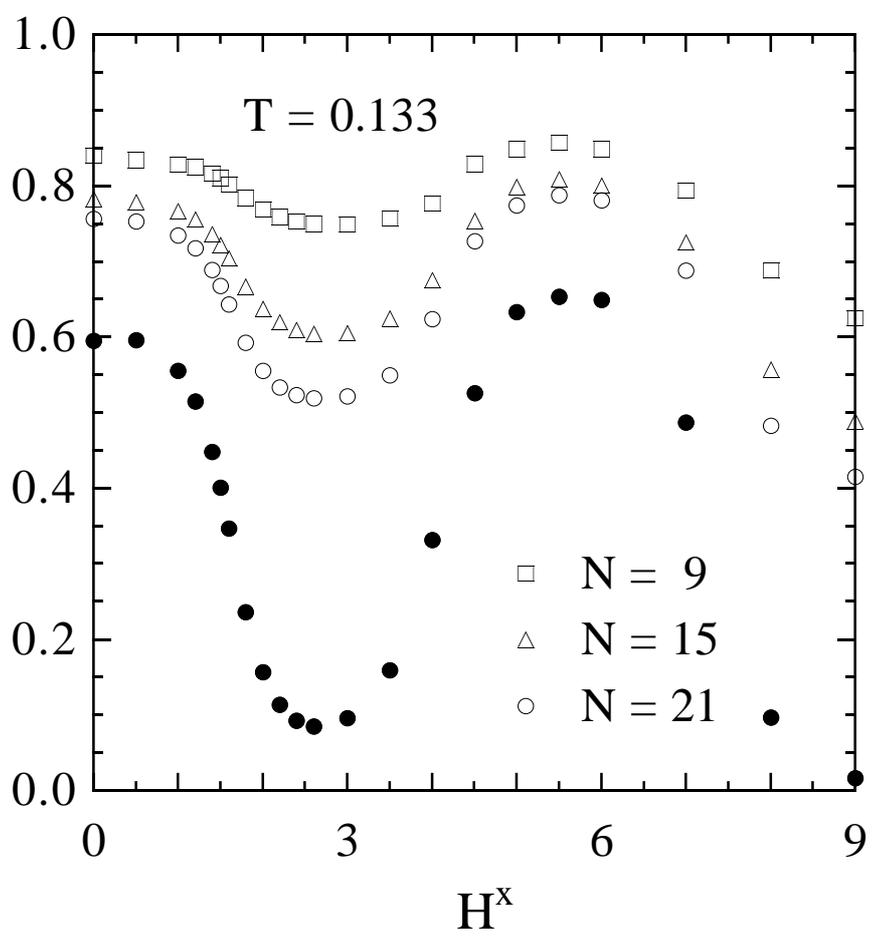
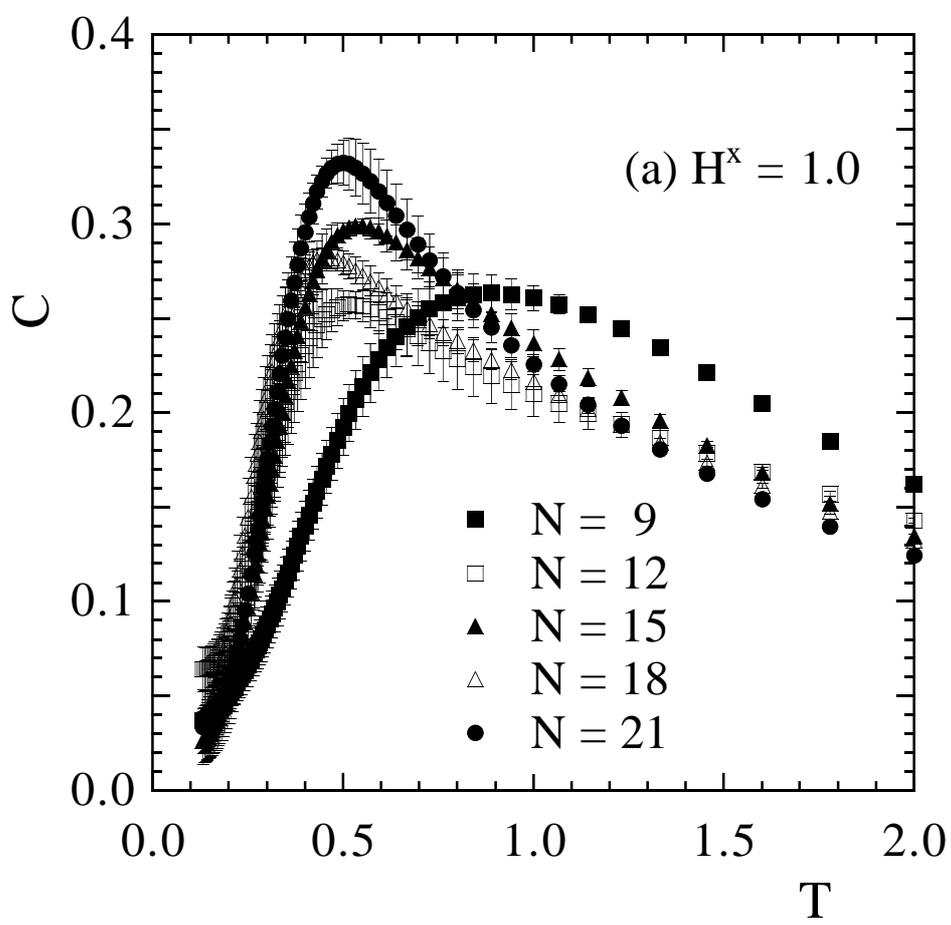
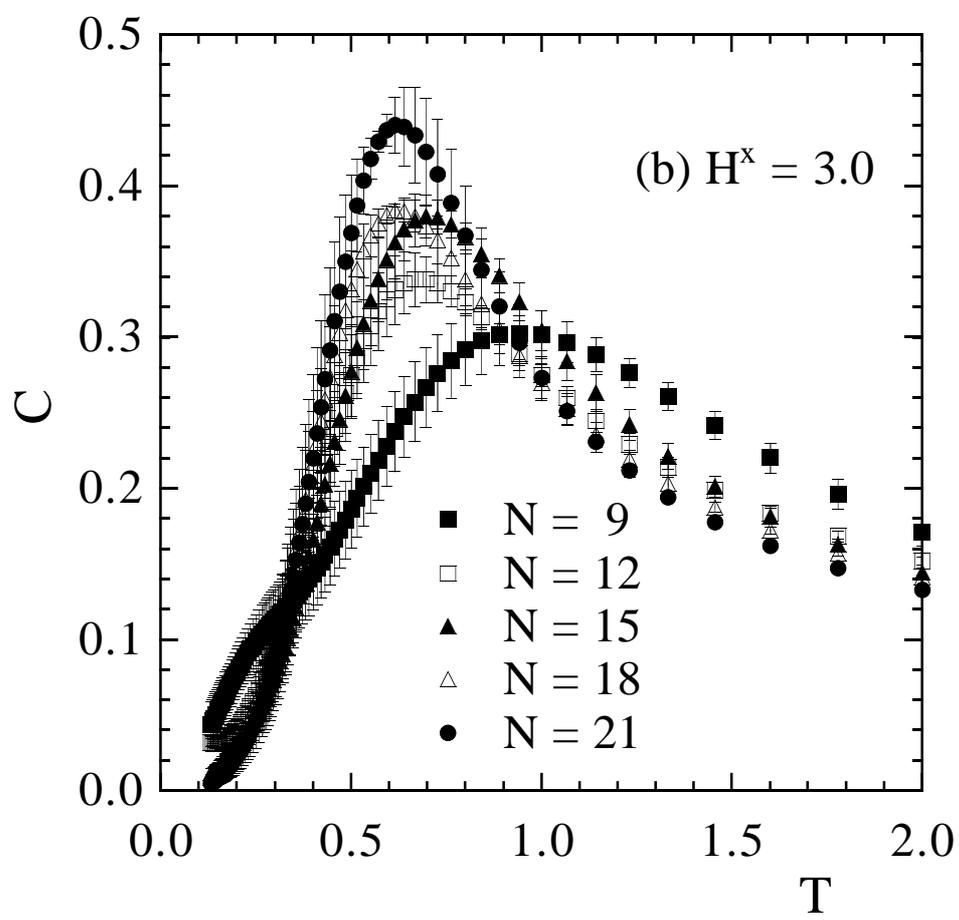
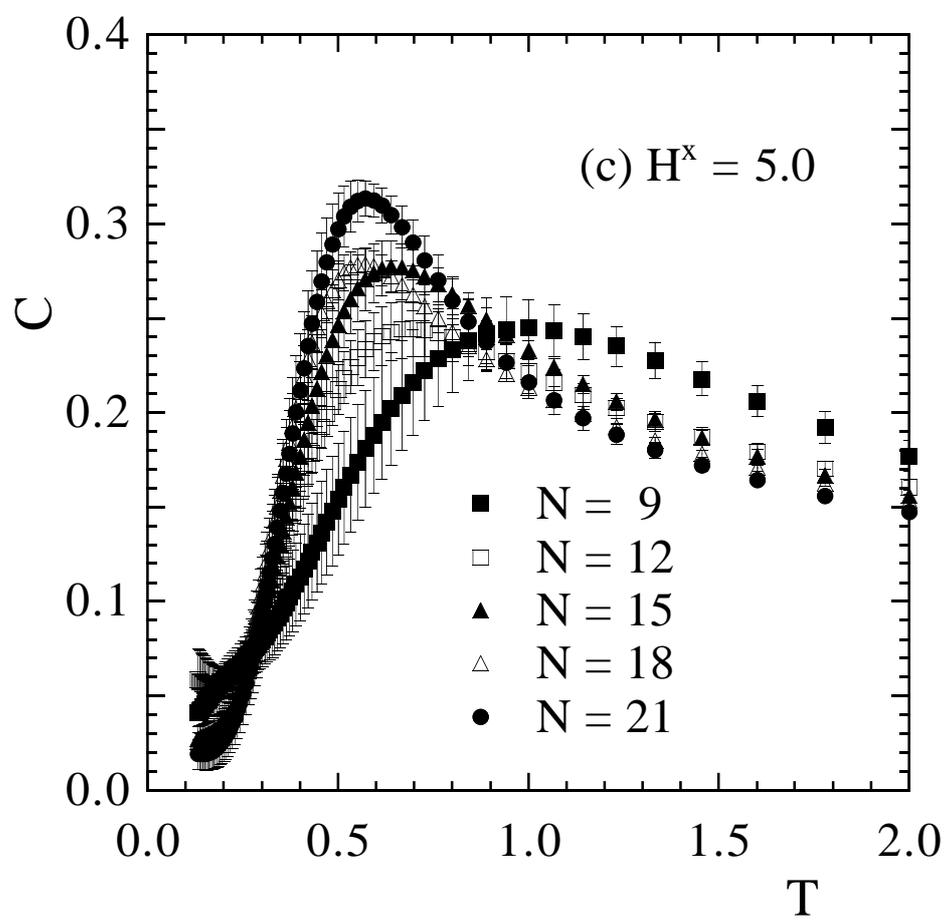
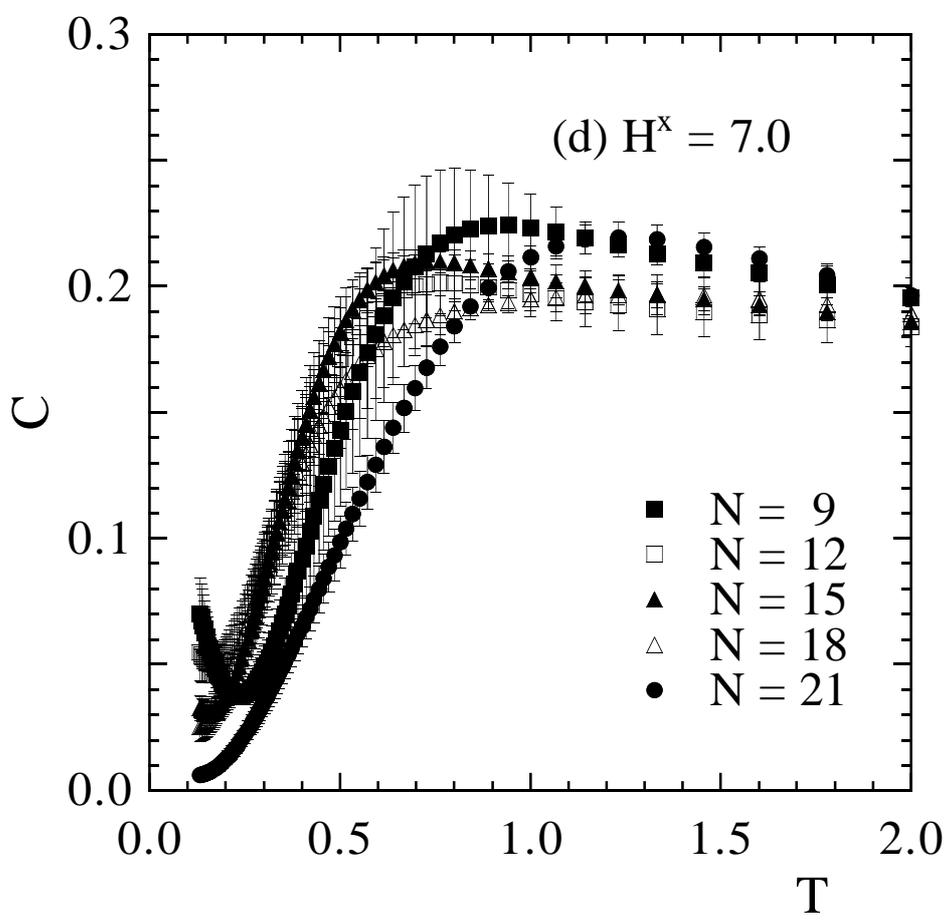
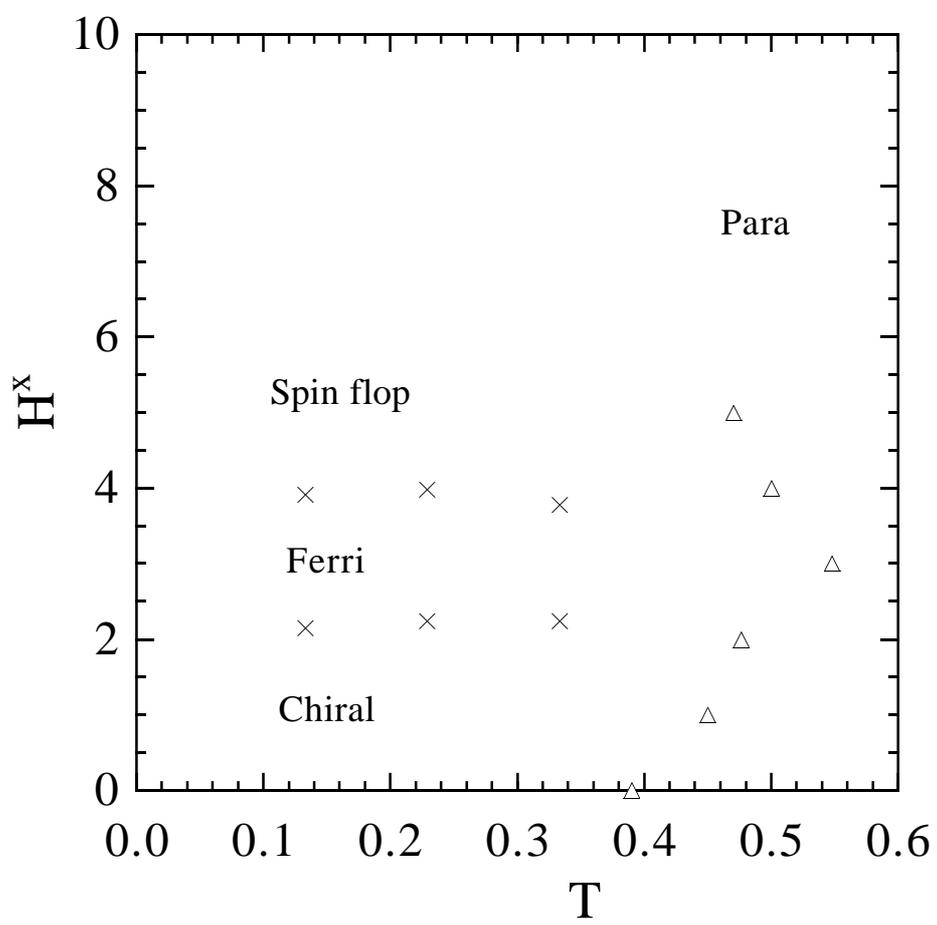
\end{document}